\documentclass[a4paper,12pt]{article}
\usepackage[utf8x]{inputenc}
\usepackage[english]{babel}
\usepackage{amsmath, amssymb, amsfonts}
\usepackage{fullpage, indentfirst, cite}
\usepackage{epsfig,graphicx}

\newcommand{\be}{\begin{equation}}
\newcommand{\ee}{\end{equation}}

\DeclareMathOperator*{\Res}{Res}
\DeclareMathOperator{\Li}{Li}
\DeclareMathOperator{\I}{I}
\DeclareMathOperator{\J}{J}

\title{
{\Large Fermions and Kaluza--Klein vacuum decay:\\ a toy model}
\date{}
\author{
  V.~A.~Rubakov$^{a,b}$\footnote{rubakov@inr.ac.ru}
  and M.~Yu.~Kuznetsov$^{a,c}$\footnote{mkuzn@inr.ac.ru}
\vspace{.2cm}\\
\footnotesize\llap{$^a$}\it Institute for Nuclear Research of the Russian Academy of Sciences, \\
\footnotesize \it 60th October Anniversary Prospect 7a, 117312 Moscow, Russia \\
\footnotesize\llap{$^b$}\it Faculty of Physics, Moscow State University, Vorobjevy Gory, 119991 Moscow, Russia \\
\footnotesize\llap{$^c$}\it Moscow Institute of Physics and Technology, \\
\footnotesize \it Institutskii per. 9, Dolgoprudny, 141700 Moscow Region, Russia}
}

\begin{document}

\maketitle

\begin{abstract}
We address the question of whether or not fermions with
twisted periodicity condition suppress
the  semiclassical decay of $M^4 \times S^1$
Kaluza--Klein vacuum. We consider a toy
(1+1)-dimensional model with twisted fermions
in cigar-shaped Euclidean background geometry
and calculate the fermion determinant. We find
that contrary to expectations, the determinant is finite.
We consider this as an indication that
twisted fermions do {\it not} stabilize
the Kaluza--Klein vacuum.
\end{abstract}


\section{Introduction}
It is known that $M^4 \times S^1$ Kaluza--Klein vacuum is unstable \cite{Witten:1981gj}.
The vacuum decay proceeds through the Euclidean bounce, whose metric is
\be
\label{bounce}
ds^2=\frac{d\xi^2}{1-\frac{R^2}{\xi^2}} + \xi^2 d\Omega^2 + R^2 \left(1-\frac{R^2}{\xi^2}\right)d\theta^2 \;,
\ee
where $R<\xi <\infty$, $0 \le \theta <2\pi$ and $d\Omega^2$ is the metric of unit 3-sphere.
At large $\xi$, the geometry of this solution is $R^4 \times S^1$, whereas as $\xi \rightarrow R$,
the geometry approaches $R^2 \times S^3$. The latter property is seen by performing the change
of variables $\xi = R + \frac{r^2}{2R}$, which gives near $r=0$
\be
\label{bounce_asym}
ds^2=dr^2+r^2d\theta^2 + R^2d\Omega^2 \;.
\ee
Upon continuation to the space-time of Minkowskian signature, the bounce \eqref{bounce}
describes the decay of $M^4\times S^1$ into nothing \cite{Witten:1981gj}.
The decay of the Kaluza--Klein vacuum and similar processes are of interest from
several viewpoints, and various versions of this phenomenon have been
extensively discussed in literature~\cite{Dowker:1995gb, Balasubramanian:2002am, Birmingham:2002st, Balasubramanian:2005bg, Aharony:2002cx,
He:2007ji, Horowitz:2007pr, Astefanesei:2005yj, BlancoPillado:2010df, BlancoPillado:2010et, Brown:2010mf, Brown:2011gt, Stotyn:2011tv}.

It has been argued in Ref.\cite{Witten:1981gj} that $M^4\times S^1$ Kaluza--Klein vacuum is stabilized in
a theory containing fermions with twisted periodicity condition in the compact coordinate,
\be
\label{angbc}
\psi(\theta)=e^{-2i\pi q}\psi(\theta+2\pi) \;.
\ee
The argument is that in the background of the bounce \eqref{bounce}, this condition makes perfect
sense away from $\xi = R$, but for generic $q$ it appears singular at $\xi = R$. The way to see the effect of fermions on
the Kaluza--Klein vacuum decay would be to calculate the ratio of fermion determinants
in the background \eqref{bounce} and in the background of the original Kaluza--Klein metric,
both with periodicity condition \eqref{angbc}:
\be
\label{det_ratio}
\Lambda(q) \equiv \frac{\det\mathcal{D}^{bounce}_q}{\det\mathcal{D}^{vacuum}_q} \; ,
\ee
where $\mathcal{D}$ is the Dirac operator and index $q$ refers to the condition \eqref{angbc}.
If this ratio vanishes, the vacuum is stabilized indeed.

In this paper we consider a toy model for this situation. Namely, instead of studying the
five-dimensional theory with the metric \eqref{bounce}, we discuss two-dimensional theory
with the metric describing a sigar-shaped space:
\be
\label{toy_bounce}
ds^2=\frac{d\xi^2}{1-\frac{R^2}{\xi^2}} + R^2 \left(1-\frac{R^2}{\xi^2}\right)d\theta^2
\ee
In other words, we disregard the $S^3$ part of the geometry. Likewise, instead of the
five-dimensional Kaluza--Klein vacuum we consider $(1+1)$-dimensional theory with the spatial
dimension compactified to a circle, whose Euclidean counterpart is described by the metric
\be
\label{vacuum}
ds^2_{vac}=d\xi^2 + R^2 d\theta^2 \; .
\ee
The metrics \eqref{toy_bounce} and \eqref{vacuum} coincide as $\xi \rightarrow \infty$,
but the geometry \eqref{toy_bounce} has a smooth end at $\xi=R$, just like in the case
of the five-dimensional bounce. We impose the periodicity condition \eqref{angbc}
on the two-dimensional fermions, which again makes sense away from $\xi=0$ but appears singular
at $\xi=R$. Thus, we argue that our toy model captures the main features of the five-dimensional
Kaluza--Klein theory, which are relevant for the vacuum decay. Our purpose is to calculate
the ratio \eqref{det_ratio} in this toy model. In our calculations we consider the interval
\be
R < \xi < T \; ,
\label{may12-12-1}
\ee
where $T$ is sent to infinity in the very end of the calculation. 
We do this for the determinants in the backgrounds of
both ``bounce'' metric \eqref{toy_bounce} and ``vacuum'' one \eqref{vacuum}.
The choice of one and the same IR cutoff $T$ in these two cases 
mimics the five-dimensional situation, where $\xi$ is unambiguously
defined as the radius of the 3-sphere both for the bounce solution and Kaluza--Klein
metric.

To simplify things further, we consider {\it massless} fermions. The motivation is that a
pathology, if any, in the fermion behavior in the background metric \eqref{toy_bounce}
would emerge from a small vicinity of the point $\xi=R$, while the short-distance properties
of fermions should not be sensitive to their mass. The advantage is that we can utilize
conformal invariance of massless two-dimensional fermions (modulo conformal anomaly, which
is local, and therefore independent of $q$). Indeed, our metric \eqref{toy_bounce},
as any other $2D$ metric, is conformally flat,
\be
ds^2=\Phi^2(r)[dr^2 + r^2 d\theta^2] \; ,
\ee
where
\begin{gather}
r= 2 \; \sqrt{\frac{\xi-R}{\xi+R}} \; e^{\frac{\xi-R}{R}}\; , \label{may10-10}\\
\Phi= R  \frac{\xi+R}{2\xi} \; e^{-\frac{\xi-R}{R}} \;.
\end{gather}
Thus, instead of calculating the fermionic determinant in the background metric \eqref{toy_bounce}
we are going to perform the calculation on a plane, still with 
the periodicity condition \eqref{angbc}.

It is worth noting that the calculation of $\det\mathcal{D}^{bounce}_q$ is equivalent to the calculation of the
determinant for conventional (periodic) massless fermions of charge $(-q)$ in the
background of an instanton in the two-dimensional Abelian Higgs model, in the limit of
vanishing instanton size. If  
there is no interaction of fermions with the scalar field, then the fermionic part
of the Lagrangian in the latter model has the form
\be
\label{lagrangian}
\mathcal{L}_\psi = i \bar{\psi}\gamma^\mu(\partial_\mu-iqeA_\mu)\psi \; ,
\ee
while in the limit of vanishing instanton size, the field of the instanton ---
Abrikosov--Nielsen--Olesen vortex \cite{Nielsen:1973cs, Abrikosov:1956sx} --- has the Aharonov--Bohm form,
$A_\mu=e^{-1}\partial_\mu \theta$. The field $\psi$ obeys the periodicity condition
$\psi(\theta+2\pi)=\psi(\theta)$, so the change of variables
$\psi \to e^{iq\theta}\psi$
reduces the problem to the calculation of the fermion determinant on $R^2$ without
the gauge field background, but with the twisted condition \eqref{angbc}.

Determinants in the instanton background are well studied in $(1+1)$ dimensional models.
For scalar and vector fields the calculation was performed, for example, in Ref.\cite{Baacke:1994bk},
and for fermions in 
Refs.\cite{Nielsen:1976hs, Nielsen:1977aw}. In Refs.\cite{Burnier:2005he, Bezrukov:2005rw}, 
the calculations were performed
for chiral fermions of half-integer charge, coupled to the scalar field. Interestingly, 
the determinants
of fermions with half-integer charge do not show 
any pathology \cite{Burnier:2005he, Bezrukov:2005rw}.

Somewhat surprisingly,
in this paper we show that similar result holds in our problem: the fermion determinant 
in the background of our toy ``bounce'' 
is finite
for arbitrary\footnote{Antiperiodic fermions ($q = \pm 1/2$) are special,
see below, and are not studied in this paper.}  $q \neq \pm 1/2$. 
We consider this as an indication that twisted fermions do not, in fact,
stabilize
the $M^4\times S^1$ Kaluza--Klein vacuum.

\section{Fermion determinants}
It is convenient to split the quantity of interest, the logarithm of the
ratio \eqref{det_ratio},
as follows:
\be
\label{ln_det_expand}
\ln \Lambda(q) =
\ln \left[\frac{\det\mathcal{D}^{bounce}_q}{\det\mathcal{D}^{bounce}_{q=0}}\right] 
 -
\ln \left[\frac{\det\mathcal{D}^{vacuum}_{q}}{\det\mathcal{D}^{vacuum}_{q=0}}\right]
+ 
\ln \left[\frac{\det\mathcal{D}^{bounce}_{q=0}}{\det\mathcal{D}^{vacuum}_{q=0}}\right] \; .
\ee
According to the above discussion, we are going to calculate $\det\mathcal{D}^{bounce}$
on a plane, and $\det\mathcal{D}^{vacuum}$ on a cylinder (\ref{vacuum}). For $q=0$, the 
fermion behavior in the background metric \eqref{toy_bounce} is manifestly healthy,
so the last term must be finite. The explicit demonstration of the latter fact is
somewhat subtle; we give the corresponding analysis in Appendix.

Let us concentrate on the first two terms in the right hand side of eq.~\eqref{ln_det_expand}.
Let us note that $\Lambda (q)$ is periodic in $q$ with period 1, so we can
consider, without loss of generality, the range $-1/2 < q \le 1/2$.
Furthermore, due to $C$-invariance, $\Lambda (q)$ is symmetric under $q \to -q$.
Therefore, it is sufficient to study the theory with $0 \le q \le 1/2$. 
We are going to perform our calculation for
\be
  0 \le q < \frac{1}{2} \; .
\label{may7-1}
\ee
The case
$q=1/2$ is subtle for reasons that will become clear later, and we leave
it for the future.

\subsection{Vacuum background}
\label{subs:vacuum}

We begin with 
the vacuum-vacuum term. We recall that 
\be
\label{vac_en_log}
\ln \left[ \det\mathcal{D}^{vacuum}_q \right] = - E_q T \; ,
\ee
where $T$ is the normalization time and
$E_q$ is the Casimir energy of fermions with twisted periodicity condition
in the $R^1 \times S^1$ theory. This Casimir energy was 
calculated in Ref.~\cite{Hetrick:1988yg}, but the resulting expression there
is somewhat implicit. So, we redo the calculation here. We write
the Casimir energy
as a sum over energies of the Dirac sea levels,
\be
\label{vac_en}
E_q = -\sum_{n=-\infty}^\infty \omega_n(q) \; ,
\ee
where $n$ is the angular spectral number and $\omega_n = |n+q|/R$.
We regularize this sum by multiplying each term by 
a factor $\exp(-\varepsilon |\omega_n(q)|)$, where $\varepsilon$ is a 
small parameter sent
to zero in the end of the calculation, and obtain 
\be
\label{dirac_sea}
E_q - E_0 = -\frac1R \left[ \sum_{n=-\infty}^\infty |n+q| e^{-\varepsilon|n+q|} 
- \sum_{n=-\infty}^\infty |n| e^{-\varepsilon|n|} \right] \; .
\ee
With our convention 
(\ref{may7-1}),
this sum is straightforwardly evaluated, and  in the limit
$\varepsilon \rightarrow 0$ we obtain
\be
\label{dirac_sea_result}
\ln \left[\frac{\det\mathcal{D}^{vacuum}_{q}}{\det\mathcal{D}^{vacuum}_{0}}\right] 
= - (E_q - E_0)T 
= - \frac TR (q^2 -q)
\ee
We have compared numerically this simple result with that of 
 Ref.~\cite{Hetrick:1988yg} and found excellent agreement.

\subsection{Bounce background}
Let us now turn to the bounce-bounce term in Eq.~\eqref{ln_det_expand},
which involves the ratio of determinants on a disc $r \le a$, where 
$a$ is the IR cutoff in terms of the coordinate $r$, cf. Eq.~\eqref{may12-12-1}.
As the operator $\mathcal{D}^{bounce}$ is anti-Hermitean, its eigenvalues are purely imaginary:
\be
\label{ln_det_bounce}
\ln \det \mathcal{D}^{bounce}
=\sum\limits_{l,m}\ln(i\lambda_{l,m}) \; ,
\ee
where $l$ and $m$ are the radial and angular
spectral numbers, 
respectively, and $\lambda_{l,m}$ are the  eigenvalues of the operator $\mathcal{D}^{bounce}$,
\be
\label{maineq}
\mathcal{D}^{bounce}\psi_{l,m}=\gamma^\mu\partial_\mu \psi_{l,m}=i\lambda_{l,m}\psi_{l,m} \; .
\ee
We take Euclidean $\gamma$-matrices and spinors in the following form:
$$\gamma^0 =
\begin{pmatrix}
0 & 1\\
1 & 0
\end{pmatrix},\qquad
\gamma^1 =
\begin{pmatrix}
0 & -i\\
i & 0
\end{pmatrix}, \qquad
\psi=
\begin{pmatrix}
\phi\\
\chi
\end{pmatrix}$$
and impose the Dirichlet boundary conditions in the radial
coordinate,
\be
\label{radbc}
\phi(a,\theta)=0 \; .
\ee
We require that the eigenfunctions $\psi_{l,m}$ be square-integrable\footnote{This
requirement ensures, as usual, that
the expansion of the fermion field in the path integral, 
$\psi = \sum_{l,m} a_{l,m} \psi_{l,m}$, 
where $a_{l,m}$ is the set of the integration variables, yields
the diagonal fermion action with finite coefficients,
$S_F = \int~d^2x~ \psi^\dagger \mathcal{D}^{bounce} \psi =
\sum_{l,m} i\lambda_{l,m} a^\dagger_{l,m}a_{l,m}  \int~d^2x ~\psi^\dagger_{l,m} \psi_{l,m}$.
Note that
the operator  $\mathcal{D}^{bounce}$ is 
indeed anti-Hermitian on square-integrable solutions obeying \eqref{radbc}, 
i.e., that the boundary terms
appearing when integrating  $(\int~d^2x~\psi_1^\dagger \mathcal{D}^{bounce} \psi_2)^*$
by parts, vanish.}.

The component $\chi$ is related to $\phi$ as follows,
\be
\label{chi}
\chi=\frac{1}{i\lambda}(\partial_0 \phi + i\partial_1 \phi) \; ,
\ee
while $\phi$ obeys
\be
\Delta\phi=-\lambda^2\phi
\ee
The eigenfunctions are
\be
\phi_{l,m} (r, \theta) = \mbox{e}^{im\theta} \phi_{l,m} (r) \; ,
\ee
where 
\be
m = n+q
\label{may12-12-2}
\ee
and $n$ is integer. $\phi_{l,m}(r)$ obeys the
Bessel equation
\be
\label{besseq}
r^2 \phi_{l,m}''+ r  \phi_{l,m}' + [\lambda_{l,m}^2 r^2 - m^2] \phi_{l,m}=0 \; .
\ee
Obviously, the eigenvalues come in pairs, $\pm \lambda_{l,m}$, where $\lambda_{l,m}>0$.
Note, that $\mathcal{D}^{bounce}$ has no zero mode:
even though for $\lambda=0$ both $\phi$ and $\chi$ can be square-integrable, it is impossible to satisfy
the boundary condition \eqref{radbc}. So, we have 
\[
\ln [\det\mathcal{D}^{bounce}] = \sum_{\lambda_{l,m} >0} \ln \lambda_{l,m}^2 \; .
\]
The square-integrable solutions are  
$\phi_{l,m} =  \J_m(r \lambda_m) =  \J_{n+q}(r \lambda_{n+q})$ for $n \ge 0$ and
$\phi_{l,m} =  \J_{-m}(r \lambda_{-m}) =  \J_{-n-q}(r \lambda_{-n-q})$ for $n \le -1$. 
It is convenient
to rewrite the latter function as $ \J_{n'-q}(r \lambda_{n'-q})$, where $n' \ge 1$;
in what follows $m_q$ denotes either $n+q$ or $n'-q$.
It is straightforward to see that for these solutions, the components $\chi$ given by
Eq.~(\ref{chi}) are square integrable as well.

Let us point out a subtlety of the antiperiodic fermions, $q=1/2$. In that case
the solutions to Eq.~(\ref{maineq}) with $n=-1$ are not square integrable:
either $\phi$ or $\chi$ behaves as $r^{-1/2}$ near the origin. We leave the analysis
of this case for the future, and proceed with the model with $0 \le q <1/2$.

\subsubsection{Sum over radial spectral numbers}

The radial spectrum of eigenvalues is determined by the following equations:
\begin{eqnarray}
\label{generator1}
\J_{n+q}(a \lambda_{l,n+q})=0 \;; \quad n \ge 0 \\
\label{generator2}
\J_{n'-q}(a \lambda_{l, n'-q})=0 \;; \quad n' \ge 1 
\end{eqnarray}
Let us evaluate the sum over radial spectral numbers $l$ for a given $n$ using
$\zeta$-function method,
\[
\sum_l \left(\ln \lambda^2_{l, m_q} - \ln \lambda^2_{l, m_0} \right)=
- \frac{d}{ds} Z_m (s)\bigg|_{s=0} \; ,
\]
where
\be
\label{rsum1}
Z_m(s) \equiv \zeta_{m_q}(s)-\zeta_{m_0}(s)=\sum_{l=1}^{\infty} \left(\lambda_{l, m_q}^{-2s} - 
\lambda_{l, m_0}^{-2s}\right)
\ee
The direct calculation of this sum is complicated by the fact that the eigenvalues 
are not known analytically. To this end, a convenient tool is the
Gelfand--Yaglom formalism \cite{Gelfand:1959nq, Dunne:2007rt, Kirsten:2003py, Kirsten:2007ev},
which was successfully applied for calculations of other functional determinants.

\begin{figure}[htb!]
\includegraphics[width=13.5cm]{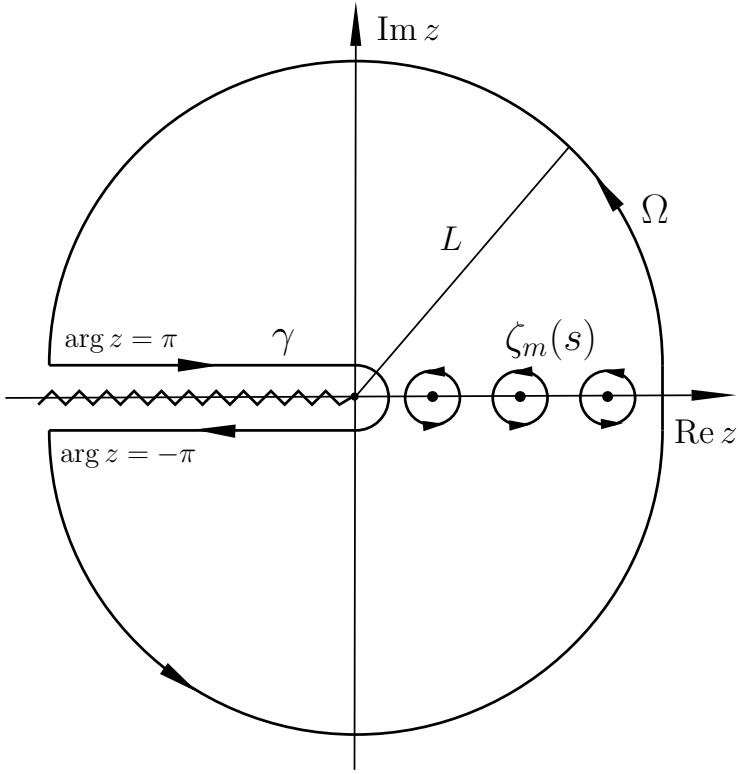}
\caption{The integration contour in the  $z$-plane. 
The sum over residues (integral over
small circles) equals the contour integral 
(\ref{may9-5}).}
\label{contour_pic}
\end{figure}

Let us briefly describe the method. 
If we know a function $F(z)$ which has zeros at desired eigenvalues, which are assumed 
to be positive, 
then its logarithmic derivative has simple poles
with residues equal to 1 at those eigenvaluse, and we can write $\zeta$-function as
follows:
\be
\zeta (s)=\sum_l \Res \left[z^{-s} \frac{d}{dz} \ln F(z)\right]
\ee
Assuming that $F(z)$ is non-singular anywhere except possibly for real
negative semi-axis and $z=\infty$, this can be transformed into a contour integral, see Fig.~\ref{contour_pic},
\be
\zeta (s) = \zeta_\gamma (s)+ \zeta_\Omega (s)  \; ,
\label{may9-5}
\ee
where
\begin{gather}
\label{intzeta1}
\zeta_\gamma (s) = \frac{1}{2\pi i}\int\limits_\gamma dz \; z^{-s}\frac{d}{dz}\ln F(z) \; , \\
\label{intzeta2}
\zeta_\Omega (s) = \frac{1}{2\pi i}\int\limits_\Omega dz \; z^{-s}\frac{d}{dz}\ln F(z) \; .
\end{gather}
The contour $\gamma$ surrounds the negative real semi-axis,
and $\Omega$ is a large circle. The integration runs clockwise in \eqref{intzeta1} and
counter-clockwise in \eqref{intzeta2}.

The first trial in our case would be $F(z) = \J_{m} (az)$,
which has zeros at $z=\lambda_{l,m}$.
However, we actually need zeros  
at $z=\lambda_{l,m}^2$. Also, we have to avoid a zero at $z=0$.
The function that satisfies  these requirements is
\be
\label{gen}
F(z)=\frac{\J_{m_q}(a\sqrt{z})}{(a\sqrt{z})^{m_q}} \; .
\ee

Let us first calculate the integral over the large circle,
\be
\label{bigcirc}
Z_{m,\Omega}(s)=\frac{1}{2\pi i}\int\limits_{\Omega} dz 
\; z^{-s}\left(\frac{d}{dz}\ln\left[\frac{\J_{m_q}(a\sqrt{z})}{\J_{m_0}(a\sqrt{z})}\right] +
\frac{d}{dz}\ln\left[a^{m_0-m_q} z^{(m_0-m_q)/2}\right]\right) \; .
\ee
To estimate the behavior of the term containing Bessel functions, we make use of the
``approximation by tangents'' \cite{grad} at large $|z|$ and possibly large index $m$, 
\begin{multline}
\J_m(z) \simeq \sqrt{\frac{2}{\pi\sqrt{z^2-m^2}}} \; \Bigg[\left(1-\frac{9}{128(z^2-m^2)}\right)\cos\left(\sqrt{z^2-m^2} - m\arccos\frac mz - \frac{\pi}{4}\right) +\\ 
\left(\frac{1}{8\sqrt{z^2-m^2}} + \frac{5m^2}{24(z^2-m^2)^{3/2}}\right)\sin\left(\sqrt{z^2-m^2}-m\arccos\frac mz - \frac{\pi}{4}\right) \Bigg],
\end{multline}
and find that the first term in the integrand
in \eqref{bigcirc}
behaves as $|z|^{-3/2}$ at complex infinity. Therefore, its contribution
to the integral \eqref{bigcirc} vanishes as $s\rightarrow 0$.
The contribution of the second term is straightforwardly evaluated and gives
\be
\label{bigcirc_res}
Z'_{m, \Omega}(0)=\frac{m_q-m_0}{2} \ln L \; ,
\ee
where $L$ is the radius of the large circle.

Let us turn to the 
remaining part,  $Z_{m , \gamma}(s)$. 
We recall that
\be
\J_m(e^{\pm i\frac{\pi}{2}}x)=e^{\pm i\frac{\pi m}{2}}\I_m(x); \quad \arg x=0 \; ,
\ee
and  get
\be
Z_{m,\gamma}(s)=\frac{\sin\pi s}{\pi}\int\limits_0^L dx \; 
x^{-s}\left(\frac{d}{dx}\ln\left[\frac{\I_{m_q}(a\sqrt{x})}{\I_{m_0}(a\sqrt{x})}\right] +
\frac{d}{dx}\ln\left[a^{m_0-m_q} x^{(m_0-m_q)/2}\right]\right)
\ee
We see that $Z^{\prime}_{m,\gamma}(0)$ is an integral of total derivative. The contribution
due to the upper limit $x=L$ is proportional to
$\ln L$ and exactly cancels out the contribution \eqref{bigcirc_res}.
Taking into account that $\ln \left[\I_{m_q}(a\sqrt{x})/\I_{m_0}(a\sqrt{x})\right]$ 
vanishes as $x\rightarrow \infty$, we finally get
\be
Z'_m(0)=-\lim_{x\rightarrow 0}\ln \left[\frac{\I_{m_q}(a\sqrt{x})}{\I_{m_0}(a\sqrt{x})}x^{(m_0-m_q)/2}\right]=
-\ln \left[\left(\frac{a}{2}\right)^{m_q-m_0} \frac{\Gamma(m_0 + 1)}{\Gamma(m_q + 1)}\right] \; .
\ee

\subsubsection{Sum over angular spectral numbers}

We are now in a position to calculate the ratio of
determinants. We regularize the sum over angular spectral
numbers by multiplying the determinants by $\exp(i\varepsilon |\partial_\theta|)$,
cf. Eq.~\eqref{dirac_sea}, and
write explicitly
\begin{multline}
\label{may12-12-3}
\ln [\det\mathcal{D}^{bounce}_q] \exp(i\varepsilon |\partial_\theta|) - \ln [\det\mathcal{D}^{bounce}_0] \exp(i\varepsilon |\partial_\theta|) \\
=\sum_{n=1}^\infty \bigg\{\ln \left[\left(\frac{a}{2}\right)^{n+q-1}\Gamma^{-1}(n+q)\right] e^{-\varepsilon (n+q-1)} +
\ln \left[\left(\frac{a}{2}\right)^{-(n-1)}\Gamma(n)\right] e^{-\varepsilon (n-1)} \bigg\} + \\
\sum_{n'=1}^\infty \bigg\{\ln \left[\left(\frac{a}{2}\right)^{n'-q}\Gamma^{-1}(n'-q+1)\right] e^{-\varepsilon (n'-q)} +
\ln \left[\left(\frac{a}{2}\right)^{-n'}\Gamma(n'+1)\right] e^{-\varepsilon n'} \bigg\}
\end{multline}
where we have shifted the argument in the first sum, $n\rightarrow(n-1)$.
The right hand side of Eq.~\eqref{may12-12-3} contains $a$-dependent and $a$-independent parts,
\be
\ln [\det\mathcal{D}^{bounce}_q] \exp(i\varepsilon |\partial_\theta|) - \ln [\det\mathcal{D}^{bounce}_0] \exp(i\varepsilon |\partial_\theta|)
=\sum_{n=1}^\infty \Delta(n,q) + \sum_{n=1}^\infty \tilde{\Delta}(n,q,a) \; ,
\ee
where
\begin{multline}
\label{reg_part_bounce}
\sum_{n=1}^\infty \Delta(n,q)= \sum_{n=1}^\infty \bigg[\ln\Gamma(n)e^{-\varepsilon n} +\ln\Gamma(n+1)e^{-\varepsilon(n+1)} \\
- \ln\Gamma(n+q)e^{\varepsilon(n+q)}-\ln\Gamma(n+1-q)e^{\varepsilon(n+1-q)}\bigg]
\end{multline}
and
\begin{multline}
\label{div_part_bounce1}
\sum_{n=1}^\infty \tilde{\Delta}(n,q,a)=\ln \frac{a}{2} \cdot
\sum_{n=1}^\infty \Big[(n+q-1)e^{-\varepsilon(n+q-1)} - (n-1)e^{-\varepsilon(n-1)}  \\
+ (n-q)e^{-\varepsilon(n-q)} - n e^{-\varepsilon n}\Big] 
\end{multline}
The $a$-dependent sum, Eq.~\eqref{div_part_bounce1},
coincides with that of Section~\ref{subs:vacuum}, and we immediately obtain that
in the limit $\varepsilon \to 0$ this term is 
\be
\label{div_part_bounce2}
\sum_{n=1}^\infty \tilde{\Delta}(n,q,a)=(q-q^2)\ln \frac{a}{2}
\ee
We now recall that $r=a$ is the cutoff radius on a plane,
which is related to the cutoff $\xi=T$ by Eq.~\eqref{may10-10}, i.e.,
$\ln (a/2) = T/R$. Thus, the contribution \eqref{div_part_bounce2}
 cancels out the second term in Eq.~\eqref{ln_det_expand},
which is given by Eq.~\eqref{dirac_sea_result}. As could have been anticipated,
the result for $\ln \Lambda (q)$ is infrared finite.

\begin{figure}
\includegraphics[width=10cm]{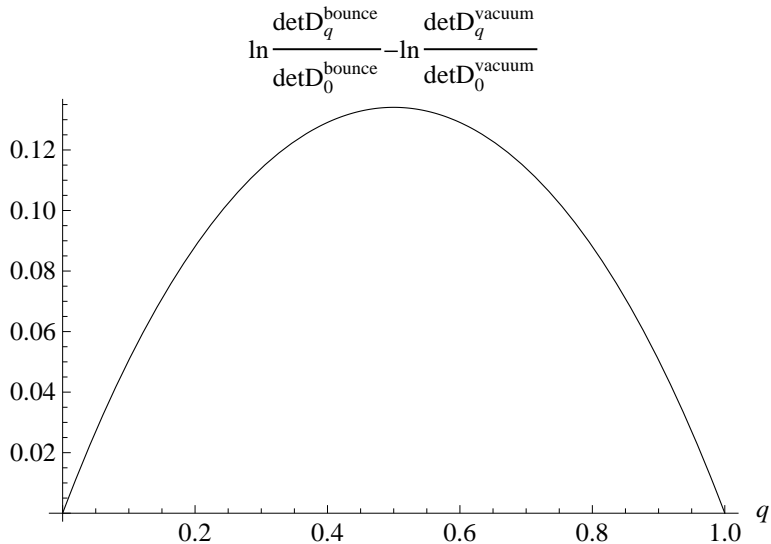}
\caption{Logarithm of the fermion determinant as function of the twist $q$.}
\label{det_pic}
\end{figure}

The quantity of interest, $\ln \Lambda (q)$, is thus given entirely by  
the $a$-independent sum \eqref{reg_part_bounce} (modulo the last,
$q$-independent term in Eq.~\eqref{ln_det_expand}). To extract its part which is
potentially
divergent in the limit $\varepsilon \to 0$, we make use of the Stirling approximation,
\[
\ln \Gamma (n) = \left[\ln \Gamma(n)\right]_{St} + O(n^{-3}) \; ,
\]
where
\be
\label{stirl}
\left[\ln \Gamma(n)\right]_{St} = n\ln n-n-\frac12 \ln n + \frac12 \ln 2\pi + \frac{1}{12n} \; .
\ee
Hereafter the subscript $St$ denotes the quantities calculated within the
Stirling approximation. Since $\Delta (n) - \Delta_{St}(n)$ decreases as $n^{-3}$ at large
$n$, potentially dangerous are the sums involving the Stirling approximation
of $\ln \Gamma(n)$. These boil down to
\begin{gather}
\sum_{n=1}^\infty e^{-\varepsilon n} n\ln n = -\frac{d}{ds} \Li_s(e^{-\varepsilon})\bigg|_{s=-1} \; ,\\
\sum_{n=1}^\infty e^{-\varepsilon n} \ln n = -\frac{d}{ds} \Li_s(e^{-\varepsilon})\bigg|_{s=0}\; ,\\
\sum_{n=1}^\infty \frac{e^{-\varepsilon n}}{n} = -\ln(1-e^{-\varepsilon}) \; ,
\end{gather}
where $\Li_s(z)$ is polylogarithm. We find that the terms that diverge as $\varepsilon \to 0$
actually cancel out due to the identity  
\be
\lim_{\varepsilon\rightarrow 0} \left(\varepsilon^2\frac{d}{ds}\Li_s(e^{-\varepsilon})\bigg|_{s=-1}
-2\varepsilon\frac{d}{ds}\Li_s(e^{-\varepsilon})\bigg|_{s=0} + \ln(1-e^{-\varepsilon})\right)=
-1-\gamma \; ,
\ee
where $\gamma$ is the Euler--Mascheroni constant.
Thus, the ratio of fermion determinants, $\Lambda (q)$ is finite, which is our main
result. The explicit expression is
\begin{multline}
\label{result}
\ln \left[\frac{\det\mathcal{D}^{bounce}_q}{\det\mathcal{D}^{bounce}_0} \right]
- \ln \left[\frac{\det\mathcal{D}^{vacuum}_{q}}{\det\mathcal{D}^{vacuum}_{0}}\right] \\
=\gamma (q-q^2) + \frac{1}{12}\left(2\gamma-1+\psi(1+q)+\psi(2-q)\right) +
 \sum_{n=1}^\infty (\Delta(n,q)-\Delta_{St}(n,q)) \; ,
\end{multline}
modulo finite and $q$-independent constant. The leading contribution here comes from the analytical part.
This function is shown in Fig.~\ref{det_pic}.

To summarize, we have found that 
all would-be divergences  cancel out, and
the determinant of twisted fermions is finite in the
background of our ``bounce''. Were similar situation inherent in the
five-dimensional theory, the Kaluza--Klein vacuum would be unstable even in
the presence of twisted fermions.

\section*{Acknowledgements}
We thank A.~A.~Belavin, A.~S.~Gorsky, S.~M.~Sibiryakov, Ye.~A.~Zenkevich, D.~V.~Kirpichnikov, A.~G.~Panin and E.~Ya.~Nugaev for
helpful discussions. We are especially indebted to P.~S.~Satunin for pointing out the Gelfand--Yaglom formalism.
This work has been supported in part by grants RFBR 12-02-00653 (V.R.), RFBR 12-02-31595 (M.K.), MK-3344.2011.2 (M.K.),
NS-5590.2012.2 and grant of Ministry of Science and Education of Russian Federation No.~8412.

\section*{Appendix}

Let us show that the last term in Eq.~\eqref{ln_det_expand}, corresponding to
periodic fermions, is finite. We begin with the vacuum term and again use Eq.~\eqref{vac_en_log},
now with $q=0$. We write
\[
E = - \frac{2}{R} \sum_{n=1}^\infty n \mbox{e}^{-\varepsilon n} \; ,
\]
where $\varepsilon$ is the regularization parameter. It is straightforward to obtain
\[
E = - \frac{2}{R\varepsilon^2} + \frac{1}{6R} \; .
\]
The interpretation of the first, divergent term is that it corresponds to the
vacuum energy density on infinite line, which should be renormalized away. Indeed,
physically meaningful is the cutoff in energy, $\mbox{exp}(-\omega_n / \Lambda)$.
Since $\omega_n = |n|/R$, we identify $\varepsilon = (\Lambda R)^{-1}$, and the first
term becomes $E= -2 \Lambda^2 R$. The corresponding energy density $E/R$ is independent of
$R$, so it is indeed the energy density on a line. So, the Casimir energy of
periodic massless fermions equals
\[
E = \frac{1}{6R} \; .
\] 

Let us consider now the bounce term. We introduce the notation
$\omega (r;R) = \ln \Phi(r;R)$, and consider the variation of the
determinant under the change of the value of $R$. The conformal anomaly
gives~\cite{Polchinski:1998rq, Belavin:2010zz}
\begin{multline}
\delta_R \ln \det\mathcal{D}_0^{bounce} = - \frac{1}{6\pi}
\int~d^2x~ \delta_R \omega \cdot \Box \omega
= \frac{1}{6\pi} \int~d^2x~\partial_\mu \delta_R \omega \cdot \partial_\mu \omega\\
- \frac{1}{6\pi} \int~d\theta~a \delta_R[\omega (r=a)] \partial_r \omega (r=a) \; .
\end{multline}
A subtlety here is that the boundary term does not vanish and, in fact, it is important
for the cancellation of the infrared divergence.
At large $r$ we have $\omega = - \ln (r/R)$, $\partial_r \omega (a) = - 1/a$ and
\[
\omega (r=a) = - \ln a + \ln R = - \frac{T}{R} + \mbox{finite} \; ,
\]
where we recalled that $\ln a/2 = T/R$ and omitted the terms which are finite
in the limit $T \to \infty$.  We immediately obtain
\[
\ln \det\mathcal{D}_0^{bounce}   
= \frac{1}{12\pi} \int~d^2x~ \partial_\mu \omega \cdot \partial_\mu \omega
- \frac{T}{3R}  + \mbox{finite} \; .
\]
This expression shows that there is no UV divergence in $\ln \det\mathcal{D}_0^{bounce}$,
as expected. 
The IR divergent part of the first term here
equals $(1/6) \ln a = T/(6R)$. Thus,
\[
\ln \det\mathcal{D}_0^{bounce} = - \frac{T}{6R}  + \mbox{finite} \; .
\]
Its IR divergent part is equal to $\ln \det\mathcal{D}_0^{vacuum} = - ET$,
so the last term   in Eq.~\eqref{ln_det_expand} is indeed finite.

\suppressfloats

\end{document}